\documentclass[iop]{emulateapj} 

\usepackage{graphicx}
\usepackage{natbib}

\def\lsim{\mathrel{\hbox{\rlap{\lower.55ex \hbox {$\sim$}}\kern-.0em
\raise.4ex \hbox{$<$}}}} 
\def\gsim{\mathrel{\hbox{\rlap{\lower.55ex \hbox {$\sim$}}\kern-.0em
\raise.4ex \hbox{$>$}}}}

\def\l{$\lambda$}

\shorttitle{Polarimety of the SLSN LSQ14mo}
\shortauthors{Leloudas et al.}

\begin{document}

\title{Polarimetry of the superluminous  supernova LSQ14mo: no evidence for significant deviations from spherical symmetry}

\author{
Giorgos Leloudas\altaffilmark{1,2}, 
Ferdinando Patat\altaffilmark{3},
Justyn~R.~Maund\altaffilmark{4},
Eric Hsiao\altaffilmark{5,6},
Daniele Malesani\altaffilmark{2},
Steve Schulze\altaffilmark{7,8},
Carlos Contreras\altaffilmark{9},
Antonio de Ugarte Postigo\altaffilmark{10,2},
Jesper Sollerman\altaffilmark{11},
Maximilian~D.~Stritzinger\altaffilmark{6},
Francesco Taddia\altaffilmark{11},
J.~Craig Wheeler\altaffilmark{12},
and Javier Gorosabel\altaffilmark{10,13,$\dagger$}  
}

\altaffiltext{1}{Department of Particle Physics and Astrophysics, Weizmann Institute of Science, Rehovot 7610001, Israel}
\altaffiltext{2}{Dark Cosmology Centre, Niels Bohr Institute, University of Copenhagen, Juliane Maries vej 30, 2100 Copenhagen, Denmark}
\altaffiltext{3}{European Southern Observatory, Karl-Schwarzschild-Strasse 2, 85748 Garching, Germany}
\altaffiltext{4}{The Department of Physics and Astronomy, University of Sheffield, Hicks Building, Hounsfield Road, Sheffield, S3 7RH, UK}
\altaffiltext{5}{Department of Physics, Florida State University, Tallahassee, FL 32306, USA}
\altaffiltext{6}{Department of Physics and Astronomy, Aarhus University, Ny Munkegade 120, 8000 Aarhus C, Denmark}
\altaffiltext{7}{Instituto de Astrof\'{\i}sica, Facultad de F\'{\i}sica, Pontificia Universidad Cat\'{o}lica de Chile, 306, Santiago 22, Chile}
\altaffiltext{8}{Millennium Institute of Astrophysics, Vicu\~{n}a Mackenna 4860, 7820436 Macul, Santiago, Chile}
\altaffiltext{9}{Las Campanas Observatory, Carnegie Observatories, Casilla 601, La Serena, Chile}
\altaffiltext{10}{Instituto de Astrof\' isica de Andaluc\' ia (IAA-CSIC), Glorieta de la Astronom\' ia s/n, E-18008, Granada, Spain}
\altaffiltext{11}{The Oskar Klein Centre, Department of Astronomy, Stockholm University, AlbaNova, 10691 Stockholm, Sweden}
\altaffiltext{12}{Department of Astronomy, University of Texas at Austin, Austin, TX 78712, USA}
\altaffiltext{13}{Unidad Asociada Grupo Ciencia Planetarias UPV/EHU-IAA/CSIC, Departamento de F\'isica Aplicada I, E.T.S. Ingenier\'ia, Universidad del Pa\'is-Vasco UPV/EHU, Alameda de Urquijo s/n, E-48013 Bilbao, Spain}

\altaffiltext{$\dagger$}{Deceased}
\begin{abstract}

We present the first polarimetric observations of a Type I superluminous supernova (SLSN).
LSQ14mo was observed with VLT/FORS2 at five different epochs in the $V$ band, with the observations starting before maximum light and spanning 26 days in the rest frame ($z=0.256$).
During this period, we do not detect any statistically significant evolution ($< 2\sigma$) in the Stokes parameters.
The average values we obtain, corrected for interstellar polarisation in the Galaxy, are $Q = -0.01\%$ $(\pm 0.15\%)$ and $U = - 0.50\%$ $(\pm 0.14\%)$.
This low polarisation can be entirely due to interstellar polarisation in the SN host galaxy.
We conclude that, at least during the period of observations and at the optical depths probed, the photosphere of LSQ14mo does not present significant asymmetries, unlike most lower-luminosity hydrogen-poor SNe Ib/c.
Alternatively, it is possible that we may have observed LSQ14mo from a special viewing angle.  
Supporting spectroscopy and photometry confirm that LSQ14mo is a typical SLSN I.
Further studies of the polarisation of Type I SLSNe are required to determine whether the low levels of polarisation are a characteristic of the entire class and to also study the implications for the proposed explosion models.

\end{abstract}




\section{Introduction}

Hydrogen-poor superluminous supernovae \citep[SLSNe;][]{2011Natur.474..487Q,2012Sci...337..927G} are rare \citep{2013MNRAS.431..912Q} stellar explosions, usually found in dwarf, metal-poor galaxies \citep{2014ApJ...787..138L,2015MNRAS.449..917L}. Despite their young environments suggesting an association with massive stars \citep{2015MNRAS.449..917L,2015MNRAS.451L..65T}, little is known about the physical mechanism driving these extraordinary transients.
The leading suggestions include powering by a central engine, such as a magnetar \citep{2010ApJ...717..245K,2010ApJ...719L.204W} or a black hole \citep{2013ApJ...772...30D}, or by interaction with circumstellar material \citep{2011ApJ...729L...6C,2012ApJ...757..178G}, possibly arising from episodic pulsational mass loss from the progenitor \citep{2007Natur.450..390W}.
Both models seem to provide adequate fits to the light curves \citep[e.g.][]{2013ApJ...773...76C} but both suffer from shortcomings and problems. 
Our lack of understanding is unsatisfactory, especially in the prospect of using SLSNe as potential distance indicators \citep{2014ApJ...796...87I} and probes of the high-$z$ Universe \citep{2012Natur.491..228C}.
As it has now been possible to collect and study large samples of H-poor SLSNe \cite[e.g.][]{2015MNRAS.452.3869N}, it is becoming clear that traditional photometry and spectroscopy alone will not be able to provide a definite answer as to what their powering mechanism is.  
Alternative and complementary methods ought to be explored.

Polarimetry is a powerful diagnostic tool to study spatially unresolved sources such as SN explosions.
A comprehensive review is given by \cite{2008ARA&A..46..433W}, who argue that core-collapse SNe are intrinsically aspherical explosions. 
This asphericity is especially prominent in the case of H-poor SNe~Ib/c \citep[e.g.][]{2007MNRAS.381..201M}, where the core of the exploding star can be directly observed. 
SNe~IIP that are surrounded by a massive H envelope show a sudden increase in the degree of polarisation after the end of the plateau phase \citep{2006Natur.440..505L,2010ApJ...713.1363C}. This can be explained by the photosphere receding within the ejecta and by the outer layers being approximately spherically symmetric. However, once the outer layers become optically thin, we are able to observe the core that has exploded asymmetrically. 
Spherical symmetry is not expected to yield a polarisation signal, as the polarisation of photons scattered in orthogonal directions cancels out. 
On the other hand, different geometries that possess a dominant axis, such as ellipsoidal or torus geometries, are expected to lead to different degrees of polarisation, and polarisation evolution, depending on the exact shape of the expanding photosphere and the observer viewing angle \citep{1991A&A...246..481H,2003ApJ...593..788K}.

In this letter we present the first polarimetric observations for an H-poor SLSN. LSQ14mo was discovered by the La Silla QUEST Survey \citep[LSQ;][]{2013PASP..125..683B} 
at coordinates $\alpha$ $=$ 10$^{\rm h}$22$^{\rm m}$41\fs53 and $\delta = -$16$^{\circ}$55$\arcmin$14\farcs4 (J2000.0). 
The Galactic extinction toward this direction is $A_V = 0.203$~mag \citep{2011ApJ...737..103S}. 
The official discovery date was on 2014 January 30 UT, but the first detection of the SN in LSQ frames dates back to January 12.2 UT.
LSQ14mo was classified during the course of the Public ESO Spectroscopic Survey of Transient Objects \citep[PESSTO;][]{2015A&A...579A..40S} by \cite{2014ATel.5839....1L} as  a SLSN I, on January 31, 04:52 UT. 
Immediately after, we triggered our dedicated program on the Very Large Telescope (VLT) in order to monitor the polarisation of a SLSN.
Observations started on February 2.2 UT.
Section~\ref{sec:obs} presents our observations and data analysis, focusing on the polarimetry. Our results are discussed in Section~\ref{sec:disc} and in Section~\ref{sec:conc} we summarise our conclusions.

\section{Observations and data analysis} \label{sec:obs}

\subsection{Spectroscopy} \label{sec:obsspec}

\begin{figure}
\plotone{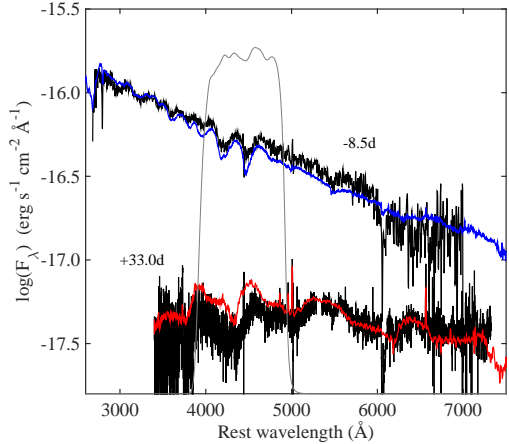}
\caption{Spectra of LSQ14mo (black) shown in the rest frame ($z = 0.256$) and offset for clarity. The rest-frame phases 
are indicated.
For comparison, smoothed spectra of PTF09cnd at $-21$d \citep{2011Natur.474..487Q} and SN~2010gx at $+30$d \citep{2010ApJ...724L..16P} are shown in blue and red, respectively.
A de-redshifted FORS2 $V$-band filter indicates which part of the SN spectrum is probed by our broadband polarimetry. \label{fig:spec}}
\end{figure}

In addition to the PESSTO classification spectrum, we secured a spectrum of LSQ14mo with the Magellan telescope equipped with IMACS on March 24, 05:40 UT. 
This spectrum covers the nominal wavelength range from 3900 to 10000\,\AA\, and consists of five individual exposures of 1200 s obtained through a $1\farcs2$-wide slit.
The classification spectrum of LSQ14mo was reduced using the PESSTO pipeline \citep{2015A&A...579A..40S}.
The Magellan spectrum was reduced in the standard manner with our own custom routines. 
The two spectra of LSQ14mo are shown in Fig.~\ref{fig:spec}.
Similar to other SLSNe I, at early phases it exhibits a characteristic blue continuum dominated by absorption features that have been identified as \ion{O}{2} \citep{2011Natur.474..487Q}, and it evolves after maximum light to resemble more regular SNe~Ic \citep{2010ApJ...724L..16P}.
Based on narrow emission lines from the host galaxy detected in the IMACS spectrum we measured a redshift of $z = 0.2561$ \citep[slightly different from the \ion{Mg}{2}  absorption redshift of 0.253 reported in][]{2014ATel.5839....1L}.
Aditionally, after correcting the spectrum for Galactic extinction, based on the Balmer decrement, we estimate that the reddening in the host galaxy explosion site is low, $E(B-V)_{\rm host} = 0.1_{-0.1}^{+0.2}$~mag.

\subsection{Photometry} \label{sec:obsphot}

\begin{figure}
\plotone{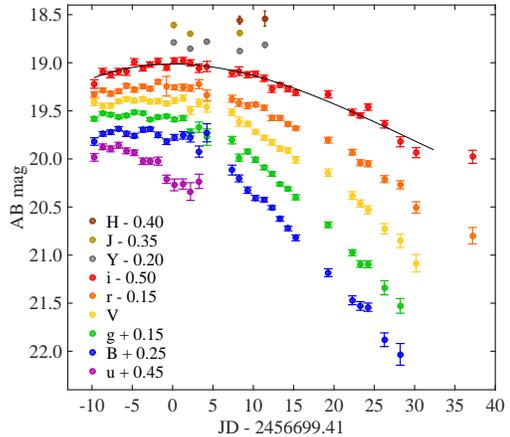}
\caption{Light curves of LSQ14mo offset for clarity as indicated. The solid line shows a third order polynomial fit to the $i$-band data. 
The time axis is with respect to the $i$-band maximum. \label{fig:LCs}}
\end{figure}

The $uBgVriYJH$ light curves of LSQ14mo were collected by the \textit{Carnegie Supernova Project-II}  \citep[CSP-II; see][Phillips et al., in preparation]{2015A&A...573A...2S}. The data were processed and the photometry was computed following \cite{2010AJ....139..519C}.
The optical and NIR photometry can be found in Tables~\ref{tab:photopt} and  \ref{tab:photnir}, and the light curves are presented in Fig.~\ref{fig:LCs}.
The CSP-II follow-up started immediately after the SN classification and close to maximum light. The time of maximum, which varies with wavelength, can be best constrained from our $i$-band data. By fitting a low-order polynomial, we estimate that maximum light occurred at $\mbox{JD} = 2456699.41$ (February~10.9 UT). The $r$ band peaked $\sim2.5$ days earlier and the bluer bands even earlier, however, the lack of sufficient pre-maximum data does not allow us to make accurate estimates. 
The $i$-band  $t_{\rm{max}}$ is the reference time that will be used throughout this paper.
By applying a k-correction based on the PESSTO spectrum, we estimate that LSQ14mo reached a maximum brightness of $M_g = -21.2$ mag.

\subsection{Polarimetry} \label{sec:obspola}

Broadband polarimetry was obtained with FORS2 on the VLT in the $V$ band.
FORS2 is a dual beam polarimeter: a turnable half-wave retarder plate (HWP) is used to rotate orthogonal polarisation components and a Wollaston prism 
separates them into an ordinary and extraordinary beam. A stripe mask is used on the focal plane  to prevent the two beams from overlapping on the detector, at the expense of reducing the effective field of view (FOV)  by half. 
Observations were obtained at four HWP angles ($0$, $22.5$, $45$ and $67.5$ deg), which allows for an optimum determination of the linear Stokes parameters.
Following \cite{2006PASP..118..146P}, the Stokes parameters $Q$ and $U$ were calculated through the normalised flux differences:

$$F_i = \frac{f_{O,i} - f_{E,i}}{f_{O,i} + f_{E,i}} $$

\noindent where $f_{O,i}$ and  $f_{E,i}$ are the intensities in the ordinary and extraordinary beams for each HWP angle $\theta_i$ respectively.
In practice, $f_{O,i}$ and  $f_{E,i}$ are the fluxes (in counts) from the SN that are measured simultaneously for each angle in a single observation.  

We obtained five epochs of broadband polarimetry spanning 26 days in the rest-frame evolution of LSQ14mo. 
A log of our observations is presented in Table~\ref{tab:pola}.
The images were reduced in a standard manner. The flat-field frames that we used were obtained without polarisation units in the light path.
To measure the SN fluxes, we used the Python package {\tt PythonPhot} \citep{2015ascl.soft01010J} to perform PSF photometry. 
Unfortunately, the number of useful PSF stars (bright and non-saturated) for each epoch was small and variable as the conditions, the exposure times, and even the exact centering of the SN (affecting the masked FOV) varied between observations. However, we were always able to construct a reliable PSF by using 2-5 stars. Photometry was obtained for a total of 22 field stars in the FOV, in addition to LSQ14mo, although their number and utility vary with the epoch of observations.
The errors in the $f_{O,i}$ and  $f_{E,i}$ fluxes were propagated to obtain the errors in the Stokes parameters $Q$, $U$, the polarisation degree $P$ and the polarisation angle $\chi$ \citep{2006PASP..118..146P}. 
The values of these parameters, describing the polarisation of LSQ14mo, can be found in Table~\ref{tab:pola}.

A consistency check was performed by observing the polarised standard star NGC~2014-1 and the unpolarised WD~0310-688 and WD~0752-676 \citep{2007ASPC..364..503F}.
The values we derive for these standards are in agreement with \cite{2007ASPC..364..503F}, confirming that  our reduction, photometry and analysis procedures are correct.

Three corrections need to be taken into account when computing $Q$, $U$, $P$, and $\chi$: 
instrumental polarisation, interstellar polarisation (ISP; polarisation induced by dust along the line of sight -- both in the Milky Way and the LSQ14mo host), and `polarisation bias'.
The instrumental polarisation caused by the instrument optics can vary across the FOV. \cite{2006PASP..118..146P} showed that FORS1 exhibited  a radial pattern in the $V$ band with polarisation increasing radially from the FOV centre.
Because the two instruments are very similar, we  
used their relation to correct the measured Stokes parameters of the field stars (the SN is not affected as it is on the optical axis).
The ISP in the Galaxy was estimated with the aid of the field stars. Their positions on the $Q$--$U$ plane were used to determine (weighted) average Galactic Stokes parameters, which were subtracted vectorially from the measured $Q$ and $U$ to obtain the intrinsic SN polarisation. 
The Galactic ISP was found to be small ($Q_{\rm{Gal}} = 0.05 \pm 0.09 \%$, $U_{\rm{Gal}} = 0.30 \pm 0.12 \%$, $P_{\rm{Gal}} = 0.30 \pm 0.12 \%$) in accordance with the maximum polarisation $P_{\rm{Gal}} < 0.6 \%$, which is expected by the low Galactic extinction toward LSQ14mo \citep{1975ApJ...196..261S,2003ARA&A..41..241D}. 
Similarly, we do not expect a large ISP in the host galaxy: the estimated $E(B-V) \sim 0.1$ mag  
translates to a limit $P_{\rm{host}} < 0.88 \%$ at the probed wavelength.
The polarisation bias is a known effect \citep[e.g.][]{1974ApJ...194..249W} related to the fact that $P$ is obtained by adding $Q$ and $U$ in quadrature 
and is therefore always a positive quantity. To correct for this effect we used the formula derived by \cite{2006PASP..118..146P} through Monte Carlo simulations  in their Section 4.1.
The $Q$, $U$, $P$ and $\chi$ for LSQ14mo that are shown in Table~\ref{tab:pola} are the values obtained after applying the corrections above.

\begin{figure}
\plotone{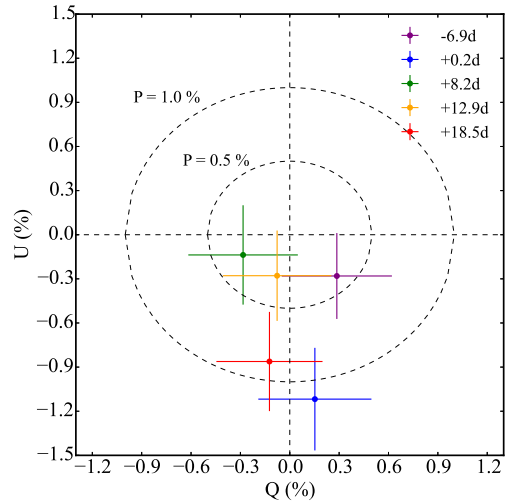}
\caption{LSQ14mo on the $Q$-$U$ plane (corrected for ISP in the Galaxy). 
The points are colour-coded according to the rest-frame phase as indicated.
Concentric circles show polarisation degrees of 0.5\% and 1.0\%.
\label{fig:QU}}
\end{figure}

Finally, we checked the stability of the Stokes parameters for the field stars during the different epochs. Since these are not expected to vary with time, from their dispersion we estimated an RMS error for each star. This RMS error was included in the ISP determination error budget but it was not applied directly to the LSQ14mo Stokes parameters in Table~\ref{tab:pola}. Stars with similar S/N as the SN have typical RMS errors of 0.38\% in Q and 0.24\% in U. This can be considered as a systematic error that dominates over the errors in Table~\ref{tab:pola}.

\section{Discussion} \label{sec:disc}

\begin{figure}
\plotone{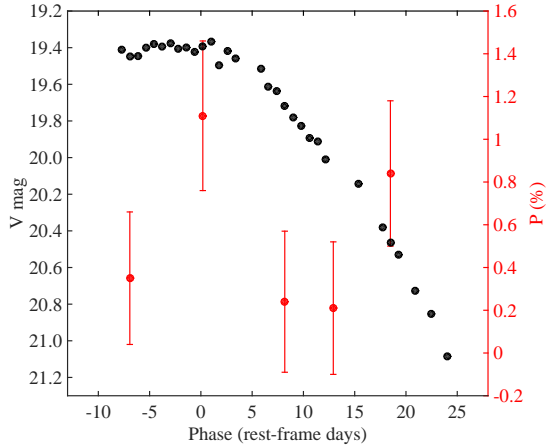}
\caption{Evolution of the $V$-band light curve of LSQ14mo (black circles; left axis; error bars have been omitted) and the polarisation measured in the $V$ band (red circles; right axis). \label{fig:PV}}
\end{figure}

Figure \ref{fig:QU} shows the location of the SN on the $Q$-$U$ plane. 
We perform a $\chi^2$ test to test the hypothesis that LSQ14mo is not polarimetrically variable, i.e. that the values of $Q$ and $U$ are constant with time.
We obtain a reduced $\chi^2$/dof of 0.46 for $Q$ and 1.63 for $U$,  around the values $Q = -0.01\%$ $(\pm 0.15\%)$ and $U = - 0.50\%$ $(\pm 0.14\%)$.
These  $\chi^2$/dof correspond to p-values of 0.76 and 0.16, respectively. There is thus no significant evidence for variability in the Stokes parameters of LSQ14mo.
The $U$ parameter demonstrates higher dispersion than $Q$ (also visible in Fig. \ref{fig:QU}), but even in this case the evidence for variability is below $2\sigma$.
Figure~\ref{fig:PV} shows the time evolution of the polarisation degree $P$ together with the $V$-band light curve. 
This is consistent with an average value of $P = 0.52\%$ $(\pm 0.15\%)$.

We only have five epochs of observations and therefore a more elaborate analysis is difficult.
However, we note the following: our second epoch is the only epoch for which the polarisation measured for the field stars 
does not agree with the radial predictions by \cite{2006PASP..118..146P} for the instrumental polarisation (but it is higher). In addition, this epoch is the main source of dispersion (RMS error) in the field stars' Stokes parameters. We were not able to understand the reasons behind this discrepancy but we note that these observations were carried relatively close to a bright Moon. Neglecting epoch 2, and by inspecting Figs.~\ref{fig:QU} and \ref{fig:PV}, it is tempting to speculate whether our last point ($+18.5$ d) tentatively indicates an increase in the polarisation. 
This would not be unprecedented: a similar behaviour is observed in SNe~IIP once the outer layers become optically thin \citep{2006Natur.440..505L,2010ApJ...713.1363C}.
The picture suggested by \cite{2010ApJ...724L..16P} in order to explain the spectroscopic evolution of SLSNe I is not very different: while we cannot explain what causes the optically thick characteristic spectra of SLSNe I before and at maximum, it is reasonable to assume that during post-maximum phases we are peering deeper in the ejecta. If the core demonstrates a higher degree of asymmetry than the outer layers we would expect the polarisation to increase. 
Even if motivated, this  analysis remains speculative and it still does not yield any statistically significant result ($p$-values $\sim 0.5$ for $Q$ and $U$). Our data do not allow us to establish any evolution in the polarisation of LSQ14mo. More and later data would be required to test this hypothesis. It was not possible to obtain such data as the fading SN quickly became too faint, making polarimetric observations very challenging.

What is certain is that we did not detect a strong polarisation signal from LSQ14mo and also that we found no significant evidence for evolution in this (weak) polarisation signal. 
The lack of polarisation evolution, which is not expected for a SN \citep{2008ARA&A..46..433W}, suggests that the remaining weak signal ($Q = -0.01\%$, $U = - 0.50\%$, $P =0.52\%$) could be explained entirely by ISP in the host of LSQ14mo. This is indeed consistent with the ISP expected from the low reddening in the host ($P_{\rm{host}} < 0.88 \%$), although this does not need to affect the SN line of sight.
However, if this polarisation is indeed due to LSQ14mo, and the accuracy of our measurements does not allow us to measure any polarisation evolution, the average $P =0.52\%$ would translate to a limit of $E > 0.9$ in the asymmetry of the photosphere \citep[assuming an oblate ellipsoid;][]{1991A&A...246..481H}.
Therefore, at least during the period of our observations, the projected photosphere of LSQ14mo appears close to spherically symmetric. 
On the other hand, \cite{2011MNRAS.415.3497D} argue that for some SN atmospheres, where the continuum and the line formation regions overlap, and the electron scattering optical depth exceeds unity, cancellation effects might lead to a suppression of the polarisation, even for significantly aspherical geometries. This might be relevant for extended SNe~IIP.
In any case, the Type I SLSN LSQ14mo appears less polarised than SN~2006aj  \citep[$P \sim 3-5\%$;][]{2006A&A...459L..33G,2007A&A...475L...1M}, a GRB-associated SN that has been suggested to be powered by a magnetar \citep{2006Natur.442.1018M}, and more similar to SN~2008D \citep[$P \sim 0.4-1\%$ in the continuum;][]{2010A&A...522A..14G,2009ApJ...705.1139M}, a more ordinary H-poor event for which spectropolarimetry suggests the presence of a stalled jet \citep{2009ApJ...705.1139M}. 

The light curves and spectra of LSQ14mo suggest that this is a typical SLSN I. 
We can obviously not constrain the explosion mechanism of SLSNe from this single observation. 
However, the lack of significant asymmetries suggest that if LSQ14mo is powered by CSM interaction, this CSM cannot be significantly asymmetric. 
The same constraint applies to a magnetar model. We note that magnetars have been observed  to show polarised emission in the radio \citep{2007MNRAS.377..107K}.
Models based on the radioactive decay of $^{56}$Ni deposited symmetrically in the inner part of the ejecta \citep{1982ApJ...253..785A} have already been excluded by previous studies of SLSNe I \citep[e.g.][]{2013ApJ...773...76C}.

Although broadband polarimetry does not probe spectral lines and the geometry of individual chemical elements in the same precision as spectropolarimetry, it is interesting to note that  our $V$-band observations  are centered exactly on the most prominent  feature of the early LSQ14mo spectrum, i.e. the W-shaped feature that has been identified as \ion{O}{2} \citep[][]{2011Natur.474..487Q} (Fig.~\ref{fig:spec}). The low polarisation in the broadband observations reduces the possibility that this feature is significantly polarised.

Our conclusions are not affected by the corrections we have applied for instrumental polarisation and ISP. Omitting these corrections, the Stokes parameters of LSQ14mo remain low ($Q = 0.04\%$, $U = -0.22\%$) and still do not show any significant ($< 1\sigma$) evidence for variability.

\section{Conclusions} \label{sec:conc}

The main objective of this project is to infer geometric information on H-poor SLSNe that will expand our understanding of their nature.
A pioneering step toward this direction was achieved by studying the broadband polarisation of the SLSN I LSQ14mo.
These are challenging observations, as a target of 20 mag requires $~1.5$ hr on the VLT  to obtain a reasonable accuracy ($\sim 0.2 \%$) in the Stokes parameters.
Furthermore, additional sources of polarisation, such as instrumental and interstellar polarisation, can increase the uncertainty budget to $\sim 0.3 \%$, which can become a nuisance, particularly if the target polarisation is intrinsically low.

We have shown that LSQ14mo cannot be significantly polarised, at least during the phases it was observed (between $-7$ and $+19$ days in the rest frame). 
This interval includes phases after maximum when  deeper layers of the ejecta are exposed.
In addition, we do not see any statistically significant  ($< 2 \sigma$)  evolution of the Stokes parameters with time. 
This suggests that the low levels of polarisation ($Q = -0.01\%$, $U = - 0.50\%$, $P =0.52\%$) can be entirely explained by ISP in the host galaxy. 
Such a low ISP is consistent with the low levels of dust extinction in the host of LSQ14mo. 
On the other hand, if the polarisation we detect is entirely due to the SN, this would imply a constraint of $E > 0.9$ in the photosphere asphericity \citep[assuming an oblate ellipsoid;][]{1991A&A...246..481H}.

The conclusion we derive is that LSQ14mo does not exhibit any significant deviation from spherical symmetry at the optical depths that we are probing.
Alternatively, it is a possibility that we may be looking at it from a special viewing angle
or that cancellation effects in an extended atmosphere may have suppressed the polarisation \citep{2011MNRAS.415.3497D}.
More observations of H-poor SLSNe are required in order to establish whether the results obtained here are general or apply only to LSQ14mo. 
Extending observations to later phases or obtaining spectral polarimetry, allowing to study the polarisation of each spectral feature separately, is especially interesting.

\begin{acknowledgements}
  
In the memory of our great friend and colleague Javier Gorosabel who started this project with us but left too early.
We thank Joe Anderson,  Abdo Campillay, Carlos Corco, Consuelo Gonz\'alez, Simon Holmbo, Nidia Morrell and Mark Phillips for observing LSQ14mo. 
The research of JRM is supported through a Royal Society University Research Fellowship.
MDS and EH gratefully acknowledge support provided by the Danish Agency for Science and Technology and Innovation.
DM is supported by the Instrument Center for Danish Astrophysics.
The research of JCW is supported in part by NSF AST-1109801.
Based on data collected under ESO program 092.D-0174(A).

\end{acknowledgements}

\bibliographystyle{apj} 


\begin{thebibliography}{43}
\expandafter\ifx\csname natexlab\endcsname\relax\def\natexlab#1{#1}\fi

\bibitem[{{Arnett}(1982)}]{1982ApJ...253..785A}
{Arnett}, W.~D. 1982, \apj, 253, 785

\bibitem[{{Baltay} {et~al.}(2013){Baltay}, {Rabinowitz}, {Hadjiyska}, {Walker},
  {Nugent}, {Coppi}, {Ellman}, {Feindt}, {McKinnon}, {Horowitz}, \&
  {Effron}}]{2013PASP..125..683B}
{Baltay}, C., {Rabinowitz}, D., {Hadjiyska}, E., {et~al.} 2013, \pasp, 125, 683

\bibitem[{{Chatzopoulos} {et~al.}(2013){Chatzopoulos}, {Wheeler}, {Vinko},
  {Horvath}, \& {Nagy}}]{2013ApJ...773...76C}
{Chatzopoulos}, E., {Wheeler}, J.~C., {Vinko}, J., {Horvath}, Z.~L., \& {Nagy},
  A. 2013, \apj, 773, 76

\bibitem[{{Chevalier} \& {Irwin}(2011)}]{2011ApJ...729L...6C}
{Chevalier}, R.~A., \& {Irwin}, C.~M. 2011, \apjl, 729, L6+

\bibitem[{{Chornock} {et~al.}(2010){Chornock}, {Filippenko}, {Li}, \&
  {Silverman}}]{2010ApJ...713.1363C}
{Chornock}, R., {Filippenko}, A.~V., {Li}, W., \& {Silverman}, J.~M. 2010,
  \apj, 713, 1363

\bibitem[{{Contreras} {et~al.}(2010){Contreras}, {Hamuy}, {Phillips},
  {Folatelli}, {Suntzeff}, {Persson}, {Stritzinger}, {Boldt}, {Gonz{\'a}lez},
  {Krzeminski}, {Morrell}, {Roth}, {Salgado}, {Jos{\'e} Maureira}, {Burns},
  {Freedman}, {Madore}, {Murphy}, {Wyatt}, {Li}, \&
  {Filippenko}}]{2010AJ....139..519C}
{Contreras}, C., {Hamuy}, M., {Phillips}, M.~M., {et~al.} 2010, \aj, 139, 519

\bibitem[{{Cooke} {et~al.}(2012){Cooke}, {Sullivan}, {Gal-Yam}, {Barton},
  {Carlberg}, {Ryan-Weber}, {Horst}, {Omori}, \&
  {D{\'{\i}}az}}]{2012Natur.491..228C}
{Cooke}, J., {Sullivan}, M., {Gal-Yam}, A., {et~al.} 2012, \nat, 491, 228

\bibitem[{{Dessart} \& {Hillier}(2011)}]{2011MNRAS.415.3497D}
{Dessart}, L., \& {Hillier}, D.~J. 2011, \mnras, 415, 3497

\bibitem[{{Dexter} \& {Kasen}(2013)}]{2013ApJ...772...30D}
{Dexter}, J., \& {Kasen}, D. 2013, \apj, 772, 30

\bibitem[{{Draine}(2003)}]{2003ARA&A..41..241D}
{Draine}, B.~T. 2003, \araa, 41, 241

\bibitem[{{Fossati} {et~al.}(2007){Fossati}, {Bagnulo}, {Mason}, \& {Landi
  Degl'Innocenti}}]{2007ASPC..364..503F}
{Fossati}, L., {Bagnulo}, S., {Mason}, E., \& {Landi Degl'Innocenti}, E. 2007,
  in Astronomical Society of the Pacific Conference Series, Vol. 364, The
  Future of Photometric, Spectrophotometric and Polarimetric Standardization,
  ed. C.~{Sterken}, 503

\bibitem[{{Gal-Yam}(2012)}]{2012Sci...337..927G}
{Gal-Yam}, A. 2012, Science, 337, 927

\bibitem[{{Ginzburg} \& {Balberg}(2012)}]{2012ApJ...757..178G}
{Ginzburg}, S., \& {Balberg}, S. 2012, \apj, 757, 178

\bibitem[{{Gorosabel} {et~al.}(2006){Gorosabel}, {Larionov}, {Castro-Tirado},
  {Guziy}, {Larionova}, {Del Olmo}, {Mart{\'{\i}}nez}, {Cepa}, {Cedr{\'e}s},
  {de Ugarte Postigo}, {Jel{\'{\i}}nek}, {Bogdanov}, \&
  {Llorente}}]{2006A&A...459L..33G}
{Gorosabel}, J., {Larionov}, V., {Castro-Tirado}, A.~J., {et~al.} 2006, \aap,
  459, L33

\bibitem[{{Gorosabel} {et~al.}(2010){Gorosabel}, {de Ugarte Postigo},
  {Castro-Tirado}, {Agudo}, {Jel{\'{\i}}nek}, {Leon}, {Augusteijn}, {Fynbo},
  {Hjorth}, {Micha{\l}owski}, {Xu}, {Ferrero}, {Kann}, {Klose}, {Rossi},
  {Madrid}, {Llorente}, {Bremer}, \& {Winters}}]{2010A&A...522A..14G}
{Gorosabel}, J., {de Ugarte Postigo}, A., {Castro-Tirado}, A.~J., {et~al.}
  2010, \aap, 522, A14

\bibitem[{{H{\"o}flich}(1991)}]{1991A&A...246..481H}
{H{\"o}flich}, P. 1991, \aap, 246, 481

\bibitem[{{Inserra} \& {Smartt}(2014)}]{2014ApJ...796...87I}
{Inserra}, C., \& {Smartt}, S.~J. 2014, \apj, 796, 87

\bibitem[{{Jones} {et~al.}(2015){Jones}, {Scolnic}, \&
  {Rodney}}]{2015ascl.soft01010J}
{Jones}, D.~O., {Scolnic}, D.~M., \& {Rodney}, S.~A. 2015, {PythonPhot: Simple
  DAOPHOT-type photometry in Python}, Astrophysics Source Code Library

\bibitem[{{Kasen} \& {Bildsten}(2010)}]{2010ApJ...717..245K}
{Kasen}, D., \& {Bildsten}, L. 2010, \apj, 717, 245

\bibitem[{{Kasen} {et~al.}(2003){Kasen}, {Nugent}, {Wang}, {Howell}, {Wheeler},
  {H{\"o}flich}, {Baade}, {Baron}, \& {Hauschildt}}]{2003ApJ...593..788K}
{Kasen}, D., {Nugent}, P., {Wang}, L., {et~al.} 2003, \apj, 593, 788

\bibitem[{{Kramer} {et~al.}(2007){Kramer}, {Stappers}, {Jessner}, {Lyne}, \&
  {Jordan}}]{2007MNRAS.377..107K}
{Kramer}, M., {Stappers}, B.~W., {Jessner}, A., {Lyne}, A.~G., \& {Jordan},
  C.~A. 2007, \mnras, 377, 107

\bibitem[{{Leloudas} {et~al.}(2014){Leloudas}, {Ergon}, {Taddia}, {Nyholm},
  {Sollerman}, {Inserra}, {Scalzo}, {Benetti}, {Pastorello}, {Smartt}, {Smith},
  {Young}, {Sullivan}, {Taubenberger}, {Valenti}, {Fraser}, {Yaron}, {Gal-Yam},
  {Manulis}, {Knapic}, {Smareglia}, {Molinaro}, {Baltay}, {Ellman},
  {Hadjiyska}, {McKinnon}, {Rabinowitz}, {Walker}, {Feindt}, {Kowalski}, \&
  {Nugent}}]{2014ATel.5839....1L}
{Leloudas}, G., {Ergon}, M., {Taddia}, F., {et~al.} 2014, The Astronomer's
  Telegram, 5839, 1

\bibitem[{{Leloudas} {et~al.}(2015){Leloudas}, {Schulze}, {Kr{\"u}hler},
  {Gorosabel}, {Christensen}, {Mehner}, {de Ugarte Postigo}, {Amor{\'{\i}}n},
  {Th{\"o}ne}, {Anderson}, {Bauer}, {Gallazzi}, {He{\l}miniak}, {Hjorth},
  {Ibar}, {Malesani}, {Morell}, {Vinko}, \& {Wheeler}}]{2015MNRAS.449..917L}
{Leloudas}, G., {Schulze}, S., {Kr{\"u}hler}, T., {et~al.} 2015, \mnras, 449,
  917

\bibitem[{{Leonard} {et~al.}(2006){Leonard}, {Filippenko}, {Ganeshalingam},
  {Serduke}, {Li}, {Swift}, {Gal-Yam}, {Foley}, {Fox}, {Park}, {Hoffman}, \&
  {Wong}}]{2006Natur.440..505L}
{Leonard}, D.~C., {Filippenko}, A.~V., {Ganeshalingam}, M., {et~al.} 2006,
  \nat, 440, 505

\bibitem[{{Lunnan} {et~al.}(2014){Lunnan}, {Chornock}, {Berger}, {Laskar},
  {Fong}, {Rest}, {Sanders}, {Challis}, {Drout}, {Foley}, {Huber}, {Kirshner},
  {Leibler}, {Marion}, {McCrum}, {Milisavljevic}, {Narayan}, {Scolnic},
  {Smartt}, {Smith}, {Soderberg}, {Tonry}, {Burgett}, {Chambers}, {Flewelling},
  {Hodapp}, {Kaiser}, {Magnier}, {Price}, \& {Wainscoat}}]{2014ApJ...787..138L}
{Lunnan}, R., {Chornock}, R., {Berger}, E., {et~al.} 2014, \apj, 787, 138

\bibitem[{{Maund} {et~al.}(2009){Maund}, {Wheeler}, {Baade}, {Patat},
  {H{\"o}flich}, {Wang}, \& {Clocchiatti}}]{2009ApJ...705.1139M}
{Maund}, J.~R., {Wheeler}, J.~C., {Baade}, D., {et~al.} 2009, \apj, 705, 1139

\bibitem[{{Maund} {et~al.}(2007{\natexlab{a}}){Maund}, {Wheeler}, {Patat},
  {Baade}, {Wang}, \& {H{\"o}flich}}]{2007A&A...475L...1M}
{Maund}, J.~R., {Wheeler}, J.~C., {Patat}, F., {et~al.} 2007{\natexlab{a}},
  \aap, 475, L1

\bibitem[{{Maund} {et~al.}(2007{\natexlab{b}}){Maund}, {Wheeler}, {Patat},
  {Baade}, {Wang}, \& {H{\"o}flich}}]{2007MNRAS.381..201M}
---. 2007{\natexlab{b}}, \mnras, 381, 201

\bibitem[{{Mazzali} {et~al.}(2006){Mazzali}, {Deng}, {Nomoto}, {Sauer}, {Pian},
  {Tominaga}, {Tanaka}, {Maeda}, \& {Filippenko}}]{2006Natur.442.1018M}
{Mazzali}, P.~A., {Deng}, J., {Nomoto}, K., {et~al.} 2006, \nat, 442, 1018

\bibitem[{{Nicholl} {et~al.}(2015){Nicholl}, {Smartt}, {Jerkstrand}, {Inserra},
  {Sim}, {Chen}, {Benetti}, {Fraser}, {Gal-Yam}, {Kankare}, {Maguire}, {Smith},
  {Sullivan}, {Valenti}, {Young}, {Baltay}, {Bauer}, {Baumont}, {Bersier},
  {Botticella}, {Childress}, {Dennefeld}, {Della Valle}, {Elias-Rosa},
  {Feindt}, {Galbany}, {Hadjiyska}, {Le Guillou}, {Leloudas}, {Mazzali},
  {McKinnon}, {Polshaw}, {Rabinowitz}, {Rostami}, {Scalzo}, {Schmidt},
  {Schulze}, {Sollerman}, {Taddia}, \& {Yuan}}]{2015MNRAS.452.3869N}
{Nicholl}, M., {Smartt}, S.~J., {Jerkstrand}, A., {et~al.} 2015, \mnras, 452,
  3869

\bibitem[{{Pastorello} {et~al.}(2010){Pastorello}, {Smartt}, {Botticella},
  {Maguire}, {Fraser}, {Smith}, {Kotak}, {Magill}, {Valenti}, {Young},
  {Gezari}, {Bresolin}, {Kudritzki}, {Howell}, {Rest}, {Metcalfe}, {Mattila},
  {Kankare}, {Huang}, {Urata}, {Burgett}, {Chambers}, {Dombeck}, {Flewelling},
  {Grav}, {Heasley}, {Hodapp}, {Kaiser}, {Luppino}, {Lupton}, {Magnier},
  {Monet}, {Morgan}, {Onaka}, {Price}, {Rhoads}, {Siegmund}, {Stubbs},
  {Sweeney}, {Tonry}, {Wainscoat}, {Waterson}, {Waters}, \&
  {Wynn-Williams}}]{2010ApJ...724L..16P}
{Pastorello}, A., {Smartt}, S.~J., {Botticella}, M.~T., {et~al.} 2010, \apjl,
  724, L16

\bibitem[{{Patat} \& {Romaniello}(2006)}]{2006PASP..118..146P}
{Patat}, F., \& {Romaniello}, M. 2006, \pasp, 118, 146

\bibitem[{{Quimby} {et~al.}(2013){Quimby}, {Yuan}, {Akerlof}, \&
  {Wheeler}}]{2013MNRAS.431..912Q}
{Quimby}, R.~M., {Yuan}, F., {Akerlof}, C., \& {Wheeler}, J.~C. 2013, \mnras,
  431, 912

\bibitem[{{Quimby} {et~al.}(2011){Quimby}, {Kulkarni}, {Kasliwal}, {Gal-Yam},
  {Arcavi}, {Sullivan}, {Nugent}, {Thomas}, {Howell}, {Nakar}, {Bildsten},
  {Theissen}, {Law}, {Dekany}, {Rahmer}, {Hale}, {Smith}, {Ofek}, {Zolkower},
  {Velur}, {Walters}, {Henning}, {Bui}, {McKenna}, {Poznanski}, {Cenko}, \&
  {Levitan}}]{2011Natur.474..487Q}
{Quimby}, R.~M., {Kulkarni}, S.~R., {Kasliwal}, M.~M., {et~al.} 2011, \nat,
  474, 487

\bibitem[{{Schlafly} \& {Finkbeiner}(2011)}]{2011ApJ...737..103S}
{Schlafly}, E.~F., \& {Finkbeiner}, D.~P. 2011, \apj, 737, 103

\bibitem[{{Serkowski} {et~al.}(1975){Serkowski}, {Mathewson}, \&
  {Ford}}]{1975ApJ...196..261S}
{Serkowski}, K., {Mathewson}, D.~S., \& {Ford}, V.~L. 1975, \apj, 196, 261

\bibitem[{{Smartt} {et~al.}(2015){Smartt}, {Valenti}, {Fraser}, {Inserra},
  {Young}, {Sullivan}, {Pastorello}, {Benetti}, {Gal-Yam}, {Knapic},
  {Molinaro}, {Smareglia}, {Smith}, {Taubenberger}, {Yaron}, {Anderson},
  {Ashall}, {Balland}, {Baltay}, {Barbarino}, {Bauer}, {Baumont}, {Bersier},
  {Blagorodnova}, {Bongard}, {Botticella}, {Bufano}, {Bulla}, {Cappellaro},
  {Campbell}, {Cellier-Holzem}, {Chen}, {Childress}, {Clocchiatti},
  {Contreras}, {Dall'Ora}, {Danziger}, {de Jaeger}, {De Cia}, {Della Valle},
  {Dennefeld}, {Elias-Rosa}, {Elman}, {Feindt}, {Fleury}, {Gall},
  {Gonzalez-Gaitan}, {Galbany}, {Morales Garoffolo}, {Greggio}, {Guillou},
  {Hachinger}, {Hadjiyska}, {Hage}, {Hillebrandt}, {Hodgkin}, {Hsiao}, {James},
  {Jerkstrand}, {Kangas}, {Kankare}, {Kotak}, {Kromer}, {Kuncarayakti},
  {Leloudas}, {Lundqvist}, {Lyman}, {Hook}, {Maguire}, {Manulis}, {Margheim},
  {Mattila}, {Maund}, {Mazzali}, {McCrum}, {McKinnon}, {Moreno-Raya},
  {Nicholl}, {Nugent}, {Pain}, {Pignata}, {Phillips}, {Polshaw}, {Pumo},
  {Rabinowitz}, {Reilly}, {Romero-Ca{\~n}izales}, {Scalzo}, {Schmidt},
  {Schulze}, {Sim}, {Sollerman}, {Taddia}, {Tartaglia}, {Terreran},
  {Tomasella}, {Turatto}, {Walker}, {Walton}, {Wyrzykowski}, {Yuan}, \&
  {Zampieri}}]{2015A&A...579A..40S}
{Smartt}, S.~J., {Valenti}, S., {Fraser}, M., {et~al.} 2015, \aap, 579, A40

\bibitem[{{Stritzinger} {et~al.}(2015){Stritzinger}, {Valenti}, {Hoeflich},
  {Baron}, {Phillips}, {Taddia}, {Foley}, {Hsiao}, {Jha}, {McCully}, {Pandya},
  {Simon}, {Benetti}, {Brown}, {Burns}, {Campillay}, {Contreras},
  {F{\"o}rster}, {Holmbo}, {Marion}, {Morrell}, \&
  {Pignata}}]{2015A&A...573A...2S}
{Stritzinger}, M.~D., {Valenti}, S., {Hoeflich}, P., {et~al.} 2015, \aap, 573,
  A2

\bibitem[{{Th{\"o}ne} {et~al.}(2015){Th{\"o}ne}, {de Ugarte Postigo},
  {Garc{\'{\i}}a-Benito}, {Leloudas}, {Schulze}, \&
  {Amor{\'{\i}}n}}]{2015MNRAS.451L..65T}
{Th{\"o}ne}, C.~C., {de Ugarte Postigo}, A., {Garc{\'{\i}}a-Benito}, R.,
  {et~al.} 2015, \mnras, 451, L65

\bibitem[{{Wang} \& {Wheeler}(2008)}]{2008ARA&A..46..433W}
{Wang}, L., \& {Wheeler}, J.~C. 2008, \araa, 46, 433

\bibitem[{{Wardle} \& {Kronberg}(1974)}]{1974ApJ...194..249W}
{Wardle}, J.~F.~C., \& {Kronberg}, P.~P. 1974, \apj, 194, 249

\bibitem[{{Woosley}(2010)}]{2010ApJ...719L.204W}
{Woosley}, S.~E. 2010, \apjl, 719, L204

\bibitem[{{Woosley} {et~al.}(2007){Woosley}, {Blinnikov}, \&
  {Heger}}]{2007Natur.450..390W}
{Woosley}, S.~E., {Blinnikov}, S., \& {Heger}, A. 2007, \nat, 450, 390

\end{thebibliography}

\begin{deluxetable}{ccccccccc}
  \tablecaption{Optical photometry of LSQ14mo \tablenotemark{a} \label{tab:photopt}}
  \tablehead{
      \colhead{UT} & 
    \colhead{JD} & 
    \colhead{Phase \tablenotemark{b}} & 
    \colhead{$u$} &
    \colhead{$g$} &
    \colhead{$r$} & 
    \colhead{$i$}  &
    \colhead{$B$} & 
    \colhead{$V$} \\
    (yy-mm-dd)     &    (days)      &     (days)      &     (mag)     &       (mag)     &      (mag)      &       (mag)     &       (mag)     &    (mag)            }
  \startdata
2014-02-01.2   &     2456689.71   &	-7.7   & 19.534 (0.039) &  19.435 (0.022) &  19.481 (0.028) &  19.721 (0.044) &  19.569 (0.039) &  19.411 (0.035)  \\ 
2014-02-02.2   &     2456690.74   &	-6.9   & 19.424 (0.030) &  19.377 (0.017) &  19.435 (0.021) &  19.590 (0.032) &  19.490 (0.024) &  19.448 (0.026)  \\ 
2014-02-03.2   &     2456691.72   &	-6.1   & 19.446 (0.030) &  19.390 (0.020) &  19.472 (0.025) &  19.624 (0.033) &  19.472 (0.023) &  19.446 (0.027)  \\ 
2014-02-04.2   &     2456692.70   &	-5.3   & 19.408 (0.031) &  19.412 (0.020) &  19.449 (0.022) &  19.593 (0.034) &  19.436 (0.022) &  19.400 (0.024)  \\ 
2014-02-05.2   &     2456693.65   &	-4.6   & 19.466 (0.031) &  19.397 (0.018) &  19.391 (0.024) &  19.593 (0.038) &  19.490 (0.026) &  19.380 (0.023)  \\ 
2014-02-06.2   &     2456694.65   &	-3.8   & 19.486 (0.029) &  19.364 (0.019) &  19.431 (0.025) &  19.491 (0.034) &  19.508 (0.023) &  19.394 (0.023)  \\ 
2014-02-07.2   &     2456695.70   &	-3.0   & 19.574 (0.034) &  19.372 (0.017) &  19.395 (0.022) &  19.556 (0.030) &  19.447 (0.025) &  19.376 (0.022)  \\ 
2014-02-08.1   &     2456696.63   &	-2.2   & 19.573 (0.032) &  19.439 (0.016) &  19.419 (0.023) &  19.519 (0.027) &  19.442 (0.022) &  19.406 (0.024)  \\ 
2014-02-09.1   &     2456697.63   &	-1.4   & 19.573 (0.042) &  19.416 (0.019) &  19.350 (0.023) &  19.484 (0.030) &  19.500 (0.028) &  19.399 (0.028)  \\ 
2014-02-10.1   &     2456698.65   &	-0.6   & 19.762 (0.043) &  19.404 (0.017) &  19.394 (0.102) &  19.550 (0.028) &  19.571 (0.029) &  19.423 (0.024)  \\ 
2014-02-11.1   &     2456699.61   &	 0.2   & 19.820 (0.061) &  19.440 (0.021) &  19.406 (0.024) &  19.476 (0.026) &  19.526 (0.030) &  19.393 (0.030)  \\ 
2014-02-12.2   &     2456700.68   &	 1.0   & 19.811 (0.053) &  19.427 (0.026) &  19.406 (0.034) &  19.475 (0.031) &  19.505 (0.039) &  19.367 (0.031)  \\ 
2014-02-13.1   &     2456701.63   &	 1.8   & 19.889 (0.092) &  19.575 (0.036) &  19.415 (0.032) &  19.502 (0.033) &  19.527 (0.047) &  19.496 (0.041)  \\ 
2014-02-14.2   &     2456702.68   &	 2.6   & 19.783 (0.074) &  19.520 (0.053) &  19.373 (0.047) &  19.557 (0.044) &  19.674 (0.058) &  19.418 (0.038)  \\ 
2014-02-15.2   &     2456703.66   &	 3.4   &    \nodata	&  19.623 (0.085) &  19.489 (0.054) &  19.539 (0.053) &  19.478 (0.094) &  19.459 (0.054)  \\ 
2014-02-18.3   &     2456706.78   &	 5.9   &    \nodata	&  19.655 (0.039) &  19.524 (0.036) &  19.611 (0.034) &  19.866 (0.051) &  19.515 (0.034)  \\ 
2014-02-19.1   &     2456707.65   &	 6.6   &    \nodata	&  19.837 (0.037) &  19.565 (0.039) &  19.588 (0.045) &  19.953 (0.036) &  19.613 (0.046)  \\ 
2014-02-20.2   &     2456708.70   &	 7.4   &    \nodata	&  19.773 (0.024) &  19.596 (0.031) &  19.625 (0.032) &  20.075 (0.033) &  19.637 (0.028)  \\ 
2014-02-21.2   &     2456709.66   &	 8.2   &    \nodata	&  19.859 (0.026) &  19.585 (0.023) &  19.619 (0.031) &  20.156 (0.027) &  19.718 (0.025)  \\ 
2014-02-22.2   &     2456710.72   &	 9.0   &    \nodata	&  19.940 (0.023) &  19.619 (0.027) &  19.662 (0.032) &  20.174 (0.023) &  19.781 (0.025)  \\ 
2014-02-23.2   &     2456711.70   &	 9.8   &    \nodata	&  20.006 (0.024) &  19.725 (0.027) &  19.772 (0.036) &  20.256 (0.021) &  19.827 (0.029)  \\ 
2014-02-24.2   &     2456712.72   &	10.6   &    \nodata	&  20.113 (0.019) &  19.726 (0.024) &  19.731 (0.024) &  20.374 (0.024) &  19.893 (0.027)  \\ 
2014-02-25.2   &     2456713.73   &	11.4   &    \nodata	&  20.162 (0.020) &  19.787 (0.022) &  19.771 (0.030) &  20.473 (0.024) &  19.912 (0.028)  \\ 
2014-02-26.2   &     2456714.69   &	12.2   &    \nodata	&  20.248 (0.028) &  19.833 (0.019) &  19.811 (0.029) &  20.571 (0.030) &  20.010 (0.035)  \\ 
2014-03-02.2   &     2456718.71   &	15.4   &    \nodata	&  20.535 (0.029) &  19.956 (0.030) &  19.828 (0.037) &  20.933 (0.041) &  20.143 (0.038)  \\ 
2014-03-05.2   &     2456721.68   &	17.7   &    \nodata	&  20.826 (0.031) &  20.080 (0.029) &  20.014 (0.032) &  21.220 (0.046) &  20.381 (0.040)  \\ 
2014-03-06.2   &     2456722.67   &	18.5   &    \nodata	&  20.943 (0.035) &  20.190 (0.031) &  20.048 (0.027) &  21.279 (0.045) &  20.464 (0.037)  \\ 
2014-03-07.1   &     2456723.62   &	19.3   &    \nodata	&  20.946 (0.036) &  20.201 (0.031) &  19.962 (0.033) &  21.292 (0.043) &  20.530 (0.042)  \\ 
2014-03-09.2   &     2456725.66   &	20.9   &    \nodata	&  21.190 (0.073) &  20.360 (0.036) &  20.140 (0.039) &  21.630 (0.071) &  20.727 (0.054)  \\ 
2014-03-11.1   &     2456727.62   &	22.5   &    \nodata	&  21.378 (0.076) &  20.419 (0.042) &  20.322 (0.052) &  21.782 (0.113) &  20.853 (0.070)  \\ 
2014-03-13.1   &     2456729.59   &	24.0   &    \nodata	&     \nodata	  &  20.654 (0.059) &  20.435 (0.052) &     \nodata	&  21.085 (0.090)  \\ 
2014-03-20.1   &     2456736.59   &	29.6   &    \nodata	&	\nodata     &  20.948 (0.085) &  20.478 (0.068) &	\nodata     &	\nodata       
\enddata
\tablenotetext{a}{Not corrected for Galactic extinction.}
\tablenotetext{b}{With respect to the $i$-band maximum and in the rest frame of LSQ14mo.}
\end{deluxetable}

\begin{deluxetable}{cccccc}
  \tablecaption{NIR photometry of LSQ14mo \tablenotemark{a} \label{tab:photnir}}
  \tablehead{
      \colhead{UT} & 
    \colhead{JD} & 
    \colhead{Phase \tablenotemark{b}} & 
    \colhead{$Y$} &
    \colhead{$J$} &
    \colhead{$H$} \\ 
     (yy-mm-dd)     &    (days)      &   (days)      &   (mag)     &       (mag)     &      (mag)                 }
  \startdata
2014-02-11.1       &  2456699.59     &    0.1	&   18.990 (0.017) &   18.959 (0.023) &      \nodata	 \\
2014-02-13.1       &  2456701.61     &    1.8	&   19.053 (0.018) &   19.051 (0.020) &      \nodata	 \\
2014-02-15.1       &  2456703.64     &    3.4	&   18.982 (0.017) &	  \nodata     &      \nodata	 \\
2014-02-19.2       &  2456707.71     &    6.6	&   19.077 (0.018) &   19.040 (0.025) &   18.957 (0.041) \\
2014-02-22.4       &  2456710.85     &    9.1	&   19.016 (0.025) &	  \nodata     &   18.941 (0.079) 
  \enddata
\tablenotetext{a}{Not corrected for Galactic extinction.}
\tablenotetext{b}{With respect to the $i$-band maximum and in the rest frame of LSQ14mo.}
\end{deluxetable}

\begin{deluxetable*}{cccccccccc}
  \tablecaption{Polarimetry of  LSQ14mo \label{tab:pola}}
\tablehead{
    \colhead{UT} & 
    \colhead{JD} & 
    \colhead{Phase \tablenotemark{a}} &
   \colhead{Exp. time \tablenotemark{b}} &
    \colhead{Seeing} &
    \colhead{SNR \tablenotemark{c}} &
    \colhead{$Q$ \tablenotemark{d}} &
    \colhead{$U$ \tablenotemark{d}} & 
    \colhead{$\chi$}  &
    \colhead{$P$ \tablenotemark{e}} \\
      (yy-mm-dd)     &   (days)      &      (days)     &       (s)     &           ($''$)   &             &    (\%)      &       (\%)     &       (deg)     &    (\%)            }
  \startdata
2014-02-02.2	&	2456690.71 & $-$6.9     &  1200 &  	0.70--0.88		&	312	&  	 0.29 (0.33)      &    $-$0.28 (0.29) &   $-$22.2 (22.4) &    0.35 (0.31)   \\  
2014-02-11.1	&	2456699.65 & $+$0.2    &    800 &  	0.65--0.83		&	297	&	 0.15 (0.34)      &    $-$1.12 (0.35) &   $-$41.1 ( 8.7)  &    1.11 (0.35)   \\ 
2014-02-21.2	&	2456709.69 & $+$8.2    &  1000 &  	0.62--0.88		&	287	&	 $-$0.28 (0.33) &    $-$0.14 (0.34) &    12.9 (30.6) &    0.24 (0.33)   \\  
2014-02-27.1	&	2456715.63 & $+$12.9  &  1200 & 	0.60--0.70		&	298	&	 $-$0.08 (0.33) &    $-$0.28 (0.31) &    37.3 (32.5) &    0.21 (0.31)  \\  
2014-03-06.1	&	2456722.62 & $+$18.5  &  1200 &  	0.80--0.98	        & 	275	&	 $-$0.12 (0.32) &    $-$0.86 (0.34) &    40.9 (10.6) &    0.84 (0.34)      
  \enddata
\tablenotetext{a}{With respect to the $i$-band maximum and in the rest frame of LSQ14mo.}
\tablenotetext{b}{Per HWP angle. The total exposure time was four times larger. All observations were divided into two consecutive equal-duration cycles (observation blocks).}
\tablenotetext{c}{Average signal to noise ratio for the SN.}
\tablenotetext{d}{Corrected for ISP in the Milky Way, as determined by field stars. The average ISP is $Q_{\rm{Gal}} = 0.05 \pm 0.09\%$ and $U_{\rm{Gal}} = 0.30 \pm 0.12 \%$.}
\tablenotetext{e}{After correcting for polarisation bias \citep{2006PASP..118..146P}.}

\end{deluxetable*}

\end{document}